\documentclass[submission,copyright,creativecommons]{eptcs}
\providecommand{\event}{MARS 2015} % Name of the event you are submitting to

\usepackage[utf8]{inputenc}

\def\eg{{e.g.},~}

\usepackage{xspace}
\usepackage[svgnames]{xcolor}
\usepackage{amsmath,amssymb}

\usepackage{wrapfig}
\usepackage{subfigure}
\usepackage{enumitem}

\usepackage{floatrow}
\newfloatcommand{capbtabbox}{table}[][\FBwidth]

\usepackage{tikz} 
\usetikzlibrary{shapes.misc}
\usetikzlibrary{calc}
\usetikzlibrary{automata}
\usetikzlibrary{backgrounds}
\usetikzlibrary{decorations.pathreplacing}
\usetikzlibrary{fadings}
\usetikzlibrary{automata,positioning,shapes.geometric,fit}

\tikzset{initial text={},
    every state/.style={circle,minimum size=.4cm,draw=blue!50,very thick,fill=blue!20},
    secret/.style={minimum size=.4cm,draw=red!50,very thick,fill=red!20,rectangle},
    node distance=1.5cm,on grid,auto,
    bend angle=65}

\def\malar{M\"alardalen University\xspace}

\def\lang{{\cal L}}

\def\uppaal{\textsc{Uppaal}\xspace}

\colorlet{vert}{green!70!black}
\colorlet{rouge}{red!70!black}
\colorlet{orange}{orange!100!black}
\colorlet{bleu}{cyan!80!white!80!black}
\colorlet{gris}{black!10!white}

\newcommand{\stack}[1]{\begin{tabular}{c} #1 \end{tabular}}

\usepackage{listings}

\definecolor{gris}{rgb}{0.3, 0.3, 0.3}

\newcounter{mynote}
\newlength\mynotewidth
\title{Timed Automata for Modelling Caches and Pipelines}
\author{Franck Cassez
\institute{Macquarie University\\ Sydney, Australia}
\email{franck.cassez@mq.edu.au}
\and
Pablo Gonz\'alez de Aledo Marug\'an
\footnote{Author gratefully acknowledges the funding from projects TEC2011-28666-C04-02, TEC2014-58036-C4-3-R and grant BES-2012-055572, awarded by the Spanish Ministry of Economy and Competitivity.}
\institute{University of Cantabria\\
Santander, Spain}
\email{\quad pabloga@teisa.unican.es}
}

\def\titlerunning{Timed Automata for Modelling Caches and Pipelines}
\def\authorrunning{F. Cassez  \& P. Gonz\'alez}

\begin{document}

\maketitle

\begin{abstract}
In this paper, we focus on modelling the timing aspects of 
binary programs running on architectures featuring caches and pipelines. 
The objective is to obtain a timed automaton model to compute tight bounds
for the worst-case execution time (WCET) of the programs using model-checking tehcniques.

\end{abstract}

\section{Introduction}

We focus on modelling the timing aspects of 
binary programs running on architectures featuring caches and pipelines. 
The objective is to obtain a model to compute tight bounds
for the worst-case execution time (WCET) of binary  programs.
The reader is referred to~\cite{wcet-survey-2008} for an exhaustive presentation of WCET computation techniques and tools. The main approach for computing the WCET is based on abstract interpretation and integer linear programming.

Modelling pipelined processors with caches was first reported in~\cite{metamoc-2009,metamoc-2010} (METAMOC).
In METAMOC the model-checker \uppaal is used to compute the WCET.  METAMOC has still some  disadvantages, the main one being the requirement of manual
loop bound annotations.
In~\cite{cassez-acsd-11,cassez-acsd-13} we 
introduced a technique to compute the WCET of binary programs in a fully automated manner without the need for loop bound annotations. The binary program is modeled as an (untimed)
finite automaton and the timing aspects of 
the hardware components like the caches and the concurrent units (stages) in the pipeline (\eg fetch, decode, execute stages) are modeled with a timed automaton
or product of simple timed automata.
The executions of the program on the hardware correspond to the possible traces
in the synchronised product of the program automaton and the NTA modelling the
hardware components.
We can then reduce the computation of the WCET to a real-time model-checking problem and
use state-of-the-art tools like \uppaal~\cite{uppaal-40-qest-behrmann-06} to compute a tight bound for the WCET.

In our previous work~\cite{cassez-acsd-11,cassez-acsd-13}, we have designed and implemented some techniques (based on program slicing) to minimise the
program automaton of a binary program and demonstrated that the model-checking approach
based on timed automata provides very accurate bounds for the WCET of the binary programs
from the \malar benchmarks~\cite{bench-malar}. 
The automata modelling the different stages of the pipeline are also very compact
and captures the essential timing features of the pipeline.  
However, the other hardware components, the caches, are modelled by very detailed generic timed automata  that 
record the full state of the caches and model the exact behaviour and content of the caches.

Model checking is known to suffer from the state explosion problem.
 For timed automata where the \emph{clocks} are part of the state space it is even more
 important to try and reduce the state space of the model to analyse while preserving enough details to faithfully model the property to be checked.
In this paper, we focus on the following problem: how to automatically compute small models for the behaviours of the caches.
We show that reducing the size of the models of the hardware components 
provides substantial benefits in the model-checking approach to compute the WCET.  

%!TEX root = mars-ws.tex

\section{Programs and hardware}
\label{sec-1}

For the sake of clarity we restrict to a  simple setting: the programs can only read/write into a fixed number
of registers (not in main memory).
A special register is the \emph{program counter}, \texttt{PC}, that points to the next instruction to be executed. 
Our programs can contain  non-deterministic choices.
As we are interested in computing a \emph{worst-case execution time} (WCET), this value should be well defined.
We impose that a program do not contain executions of (finite) but arbitrarily large lengths.
In other words, the computation tree of a program  $P$ must be a bounded depth.
Although this assumptions are not realistic for general programs and architectures, previous work has already addressed the case \cite{wcet-corr-11}, and in this work we focus on how to compute small cache models
rather than design these models by hand.

A program \emph{run} (or execution) is completely defined by a finite sequence of (positive) integers given by the successive values of the program counter (there is no input data).
The execution time of a run is defined by the time it takes for the \emph{hardware} to execute the sequence
of instructions in the run.
The hardware is composed of an \emph{instruction cache} and the \emph{main memory} to store the program, and a \emph{one-stage pipeline} (CPU) to execute the instructions.

To execute a run, say $\sigma_1 = 1 . 2 . 3 . 1$, the code of instruction $1$ (\eg an addition between two registers) has to be fed to the CPU (pipeline) to be executed. The same applies for instruction $2$ and so on.
Initially, the code of each instruction  is stored in the main memory 
and the cache is empty.
To execute the instruction at \texttt{PC}$=i$ the following steps occur:
\begin{enumerate}
   \item the CPU requests the code of the instruction $i$.
   \item  If the code of $i$ is in the cache, this is  a cache \emph{Hit}, and the code of
   $i$ is transferred from the cache to the CPU, otherwise, a cache \emph{Miss} occurs. 
     The code of $i$ is fetched from main memory, stored in  the cache and then transferred to the CPU.
   \item the CPU executes the code of $i$ and is ready to process the next instruction.
 \end{enumerate} 
The cache is a fast memory component but has limited capacity (size). 
Assume our instruction cache has capacity $3$, and is initially empty. Executing  $\sigma_1 = 1. 2 . 3. 1$ will result in $3$ cache misses and a cache hit (instruction $1$).
If the cache size is $2$, and we are about to execute $3$, the cache is full. We need to remove one instruction stored in the cache, either $1$ or $2$. A standard \emph{replacement} policy is the FIFO policy where the least recently accessed item in the cache is removed. In our case, this would mean that $1$ is removed and replaced by $3$, and $2$ becomes the least recently accessed item.

In real caches, each line of the cache may contain a \emph{set} of instructions\footnote{When transferring from
the main memory to the cache, it is usually faster to transfer a sequence of consecutive instructions rather than a single one.}.
In this case, checking whether instruction $i$ is in the cache reduces to checking whether $set(i)$ is the cache.
Finally, executing the code of an instruction $i$ takes some CPU cycles  $dur(i)$.

In the next section, we show how to model the timing aspects of the hardware and how to compute the WCET of a given program.

\section{Timed automata model}

\begin{figure}[thbtp]
\subfigure[Program automaton]{
  \includegraphics[scale=0.55]{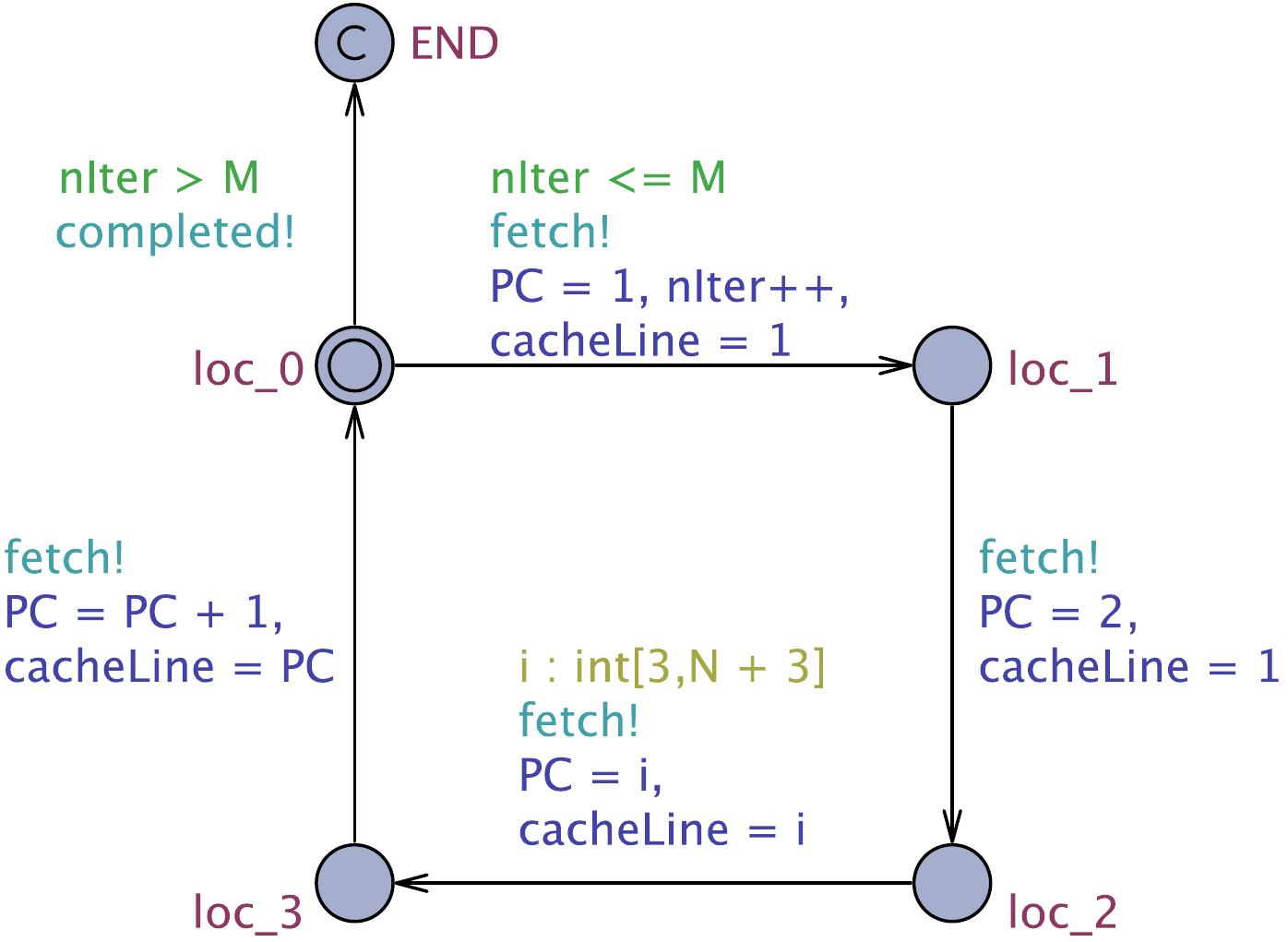}
  \label{fig-prog-auto}
}
\subfigure[CPU (two-stage pipeline, Fetch/Execute)] {
  \includegraphics[scale=0.55]{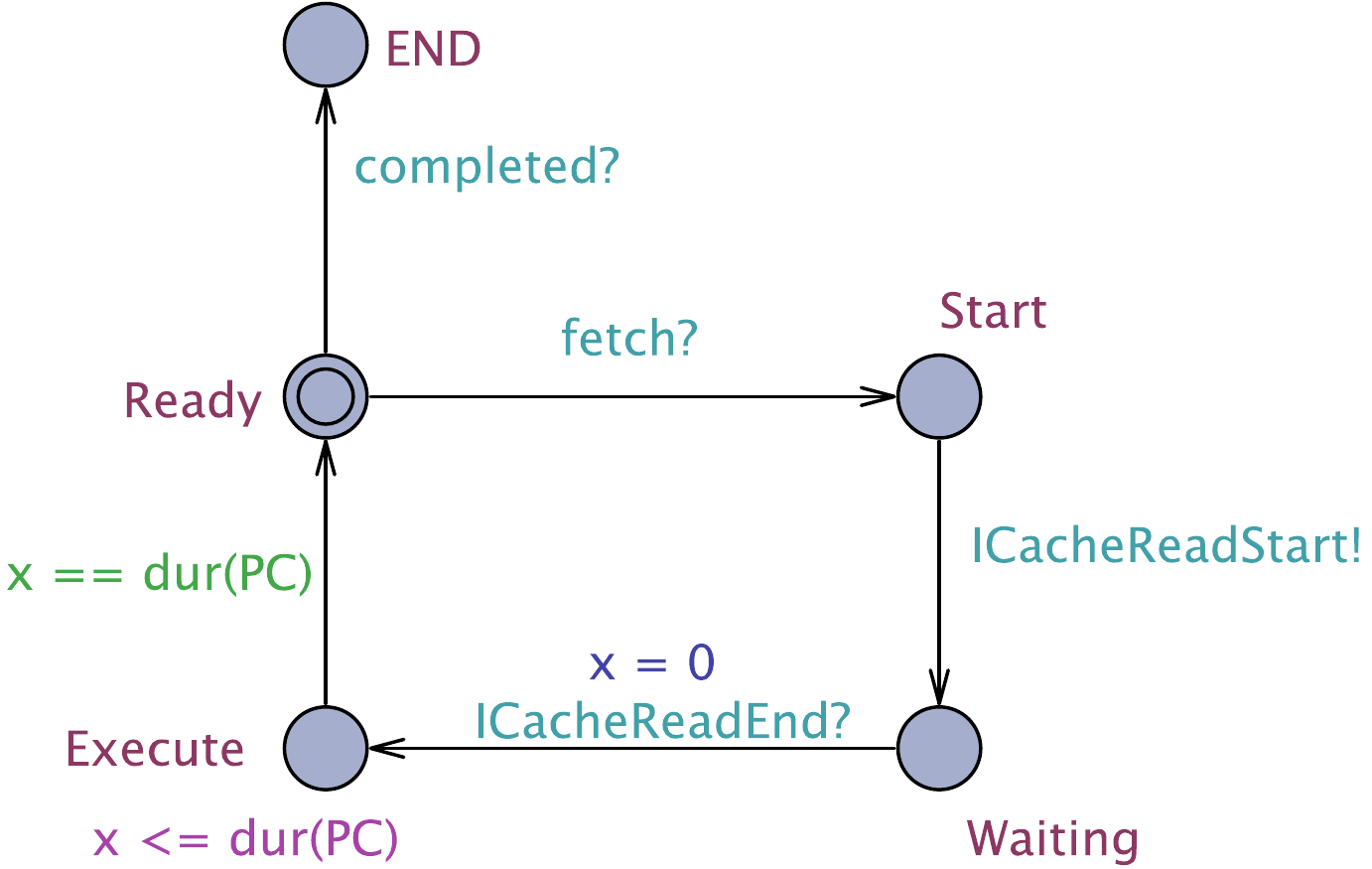}
    \label{fig-cpu-auto}
}
\caption{Program and pipeline}
\label{fig-simple1}
\end{figure}

\paragraph{\it \bfseries Program and pipeline models.}
Fig.~\ref{fig-simple1} provides an overview of our model specified as a NTA
(in \uppaal) and this will serve as a running example to illustrate our modelling techniques.
% \FC{Say what C (committed means)}
%
The program automaton (Fig.~\ref{fig-prog-auto}) models a program that performs $M$ iterations
of a loop, the body of which consists of $N$ switch statements.
The \uppaal model of Fig.~\ref{fig-prog-auto} encodes the switches as  non-deterministic choices
from location \texttt{loc\_2}.
An integer $i$ is non-deterministically picked within the interval $[3..3+N]$ and the program counter is set to this value. 
If $N=0$ there is no switch and the program has a single run. If $N=1$ there are two sets of runs (modelling an \texttt{if then else} statement): in \texttt{loc\_2}, if $i=3$ if picked, the next two instructions are $3,4$ otherwise $i=4$ and the next instructions are $4,5$.
By incrementing $N$ we can model a control loop with $N$ different inner branches. 
The variable \texttt{cacheLine} corresponds to $set(\texttt{PC})$.

We model the execution of an instruction (described in Section~\ref{sec-1}) as follows:
\begin{enumerate}
   \item the program is eager to be executed and wants to send a \texttt{fetch!} event as soon as 
  the receiver (CPU) is ready to receive (\texttt{fetch?}) it (handshake synchronisation). This is modelled in \uppaal by an \emph{urgent channel}, \texttt{fetch}. 
  \item when the \texttt{fetch!} occurs, both \texttt{PC} and \texttt{cacheLine} are set to hold the values of the next instruction to be executed and the corresponding cache set.
  \item the CPU  requests the code of the instruction by sending a \texttt{ICacheReadStart!} to the cache. This is an urgent channel as well (handshake synchronisation must happen as soon as both the sender and the receiver are ready).
  \item When the instruction is in the CPU, the cache issues the \texttt{ICacheReadEnd!} event.
  \item the CPU  receives this event \texttt{ICacheReadEnd?} and is ready to execute the code
  of the instruction. Executing the instruction  takes $dur(\mathtt{PC})$. We encode the duration by using a \emph{clock} $x$. The clock  $x$  is reset when entering the \texttt{Execute} location and  the invariant $x \leq dur(\mathtt{PC})$ and the guard $x == dur(\mathtt{PC})$ (from \texttt{Execute} to \texttt{Ready}) ensures that the automaton remains $dur(\mathtt{PC})$
  cycles in \texttt{Execute} before moving to \texttt{Ready}.
\end{enumerate}

\begin{wrapfigure}{r}{0.55\textwidth}
  \vspace{-5pt}
  \begin{center}
   \includegraphics[trim=0 20 6 15,clip,width=.85\textwidth]{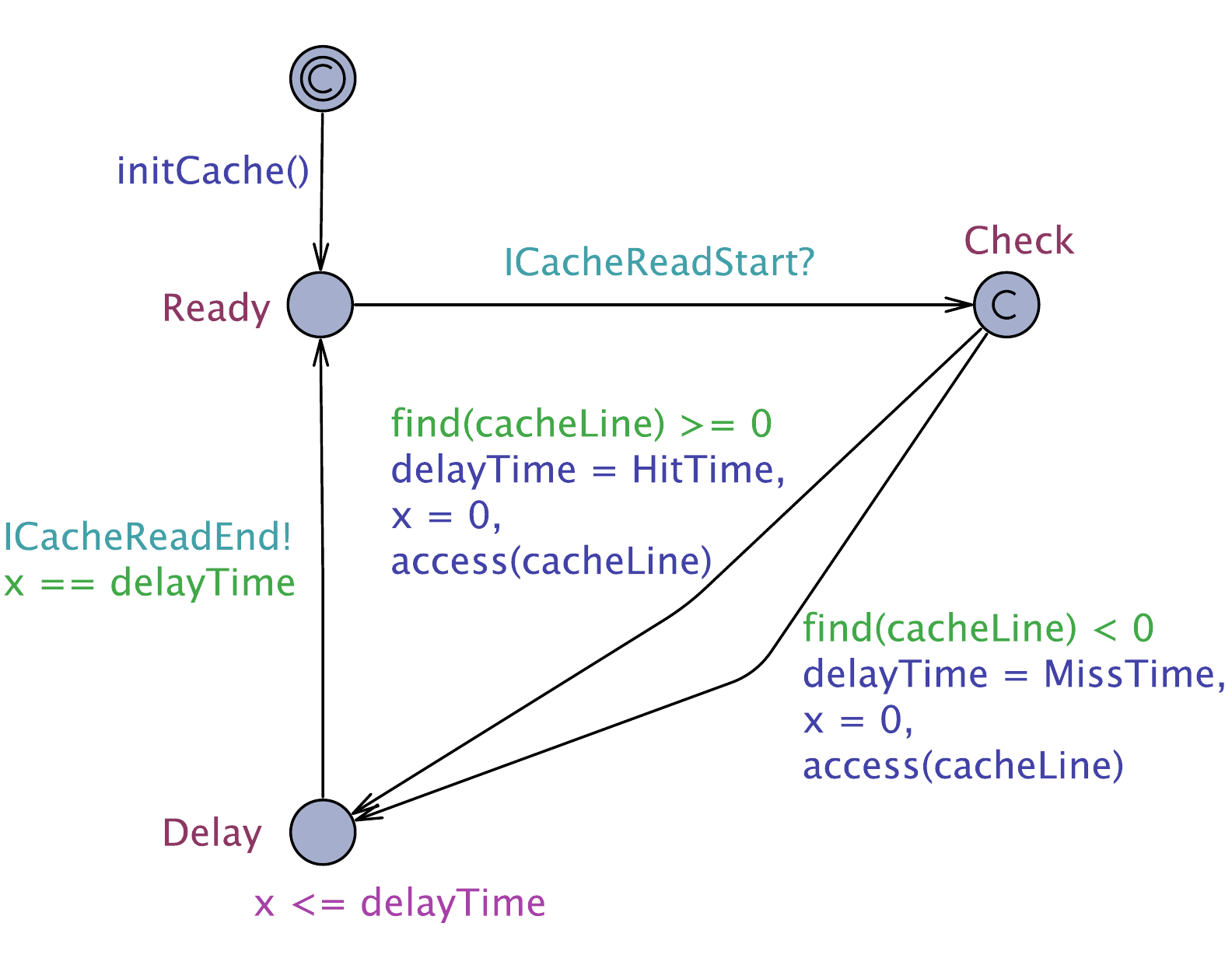}
  \end{center}
  \vspace{-10pt}
  \caption{Instruction Cache automaton \texttt{FullCache}}
  \label{fig-fullcache}
  % \vspace{-10pt}
\end{wrapfigure}

\paragraph{\it \bfseries Cache model.}
The model for the instruction cache is given by the timed automaton \texttt{FullCache} Fig.~\ref{fig-fullcache}.
When a cache read request is received (\texttt{ICacheReadStart?}) location \texttt{Check} is reached.
\noindent The location \texttt{Check} is \emph{committed} which in \uppaal implies that time cannot elapse and we must leave this location
instantaneously\footnote{This is a loose definition of the actual  semantics of committed in \uppaal but valid for our model.}.

If the instruction \texttt{PC} is in the cache, \texttt{find(cacheLine)} returns a positive value, otherwise
a negative value.
The timing behaviour of the cache depends on whether the instruction \texttt{PC} is in the cache or not:
if \texttt{PC} is in the cache, a \emph{Hit} occurs and the delay is set to \texttt{HitTime} (\eg $2$ cycles). 
Otherwise, a \emph{Miss} occurs and the delay is set to \texttt{MissTime} (\eg $20$ cycles).
In both cases the cache is updated by executing the \texttt{access(cacheLine)} function.
 % and the cache is updated (\texttt{access(cacheLine)}).
The \uppaal specification of the  functions \texttt{find($\cdot$)} and \texttt{access($\cdot$)} are given in Appendix~\ref{app-cache-full}.

%!TEX root = mars-ws.tex

\section{Computing the WCET}
\label{sec-mc}

\paragraph{\it \bfseries Real-time model checking.}
To compute the WCET of our program, we synchronise the previous components, program, CPU (pipeline) and 
cache to build an NTA.
Assume we have a clock \texttt{GBL\_CLK} which is initially set to $0$ and never reset.
The location \texttt{END} of the program (Fig.~\ref{fig-prog-auto}) is committed and has no outgoing edges; this implies that time cannot elapse
and no other transition can be taken, so it creates a deadlock. This ensures that when we reach the end of the program,
the value of \texttt{GBL\_CLK} is frozen. Without blocking the system that way,  \texttt{GBL\_CLK} could grow
arbitrarily large in the \texttt{END} location of the program. 
In this model, the WCET of our program running on the simple hardware, is the maximal value that \texttt{GBL\_CLK} can hold at the end of the execution of the program, in location \texttt{END}.
This can be easily computed in \uppaal with the \emph{sup} operator.

In our previous~\cite{cassez-acsd-11,cassez-acsd-13,wcet-corr-11} work we designed a very accurate model of the ARM920 hardware and compared the results obtained with our method (based on timed automata), with actual execution times on a real ARM board. On the set of benchmarks from \malar we managed to compute very tight WCETs.

\paragraph{\it \bfseries Advantages of timed automata modelling.}
Using timed automata to model (binary) programs running on complex hardware 
has several advantages:
\begin{itemize}
  \item NTA  have precise semantics; our models are formal behavioural models for the hardware.
  \item we define the model of a program running on some hardware as the composition (synchronisation) of very simple components. All these components can be validated separately.
  \item we can model features that cannot be modelled with other approaches based on abstract interpretation and Integer Linear Programming (ILP); for example, if the execution time of an instruction is within an interval (depends on the data to be processed), we can easily model the duration by an interval in the CPU model.
  We can also model  changes of the speed of the CPU during program executions. 
  \item our models and methodology ensures that we obtain an upper bound of the actual WCET provided that the models of the hardware are certified to be good abstractions (time-wise) of the actual board.
  \item our methodology is resilient to \emph{timing anomalies}~\cite{wcet-12}. This is a challenging problem faced by other methods based on abstract interpretation. 
\end{itemize}
On the other hand, model-checking is very sensitive to the number of states of the system to analyse: this is the well known state explosion problem that can hinder the analysis of large systems.
This is even more true when model-checking timed automata:
the state space is composed of a discrete part and clock constraints.
 
\paragraph{\it \bfseries Reducing the state space.}
The model checking approach we have introduced is mostly state based: the program and the hardware are modelled as timed automata and we can generate all the possible states of the system and obtain the maximal value of a clock in a particular location.

A key concept in our approach~\cite{cassez-acsd-11} is to define the WCET problem as a language based problem:
\begin{itemize}
  \item the program $P$ is viewed as a generator of sequences of instructions; the sequences are in a language
    $\lang(P)$ generated by an automaton $Aut(P)$.
  \item the hardware $H$ is a transducer: given a  sequence $w \in \lang(P)$, it computes an integer value, the execution time of $w$,  number of  cycles.
\end{itemize}
The information needed to compute the WCET is captured by the sequences of instructions (the \texttt{PC}). We do not need to simulate the actual effect of each instruction.
In other words, if we can generate $\lang(P)$ and we have a timed automaton model of the hardware, we can compute the WCET by examining all the words in $\lang(P)$ and ``running'' them in the model of the hardware (viewed as a transducer).
If we generate $\lang(P)$ with the actual program by keeping track of the full state of the program (registers, stack), we may end up generating the same sequence of instructions many times.

In~\cite{cassez-acsd-11,cassez-acsd-13,wcet-corr-11} we have 
shown  how to compute (using program slicing) a reduced program $P'$ and an associated program automaton $Aut(P')$, 
such that $\lang(P') = \lang(P)$ but $Aut(P')$ is \emph{minimal}: it generates every word in $\lang(P)$ only once.
We have demonstrated that this is essential for being able to compute the WCET of real programs.
In practice, using $Aut(P')$ avoids feeding the hardware transducer twice with the same sequence of instructions hence reducing the amount of work of the real-time model-checker.
Still, in our approach the caches are modelled explicitly and this is not optimal as demonstrated in the next paragraph.

\begin{figure}[hbtp]
\begin{floatrow}
\ffigbox{%
   \includegraphics[trim=90 0 0 0,clip,width=.63\textwidth]{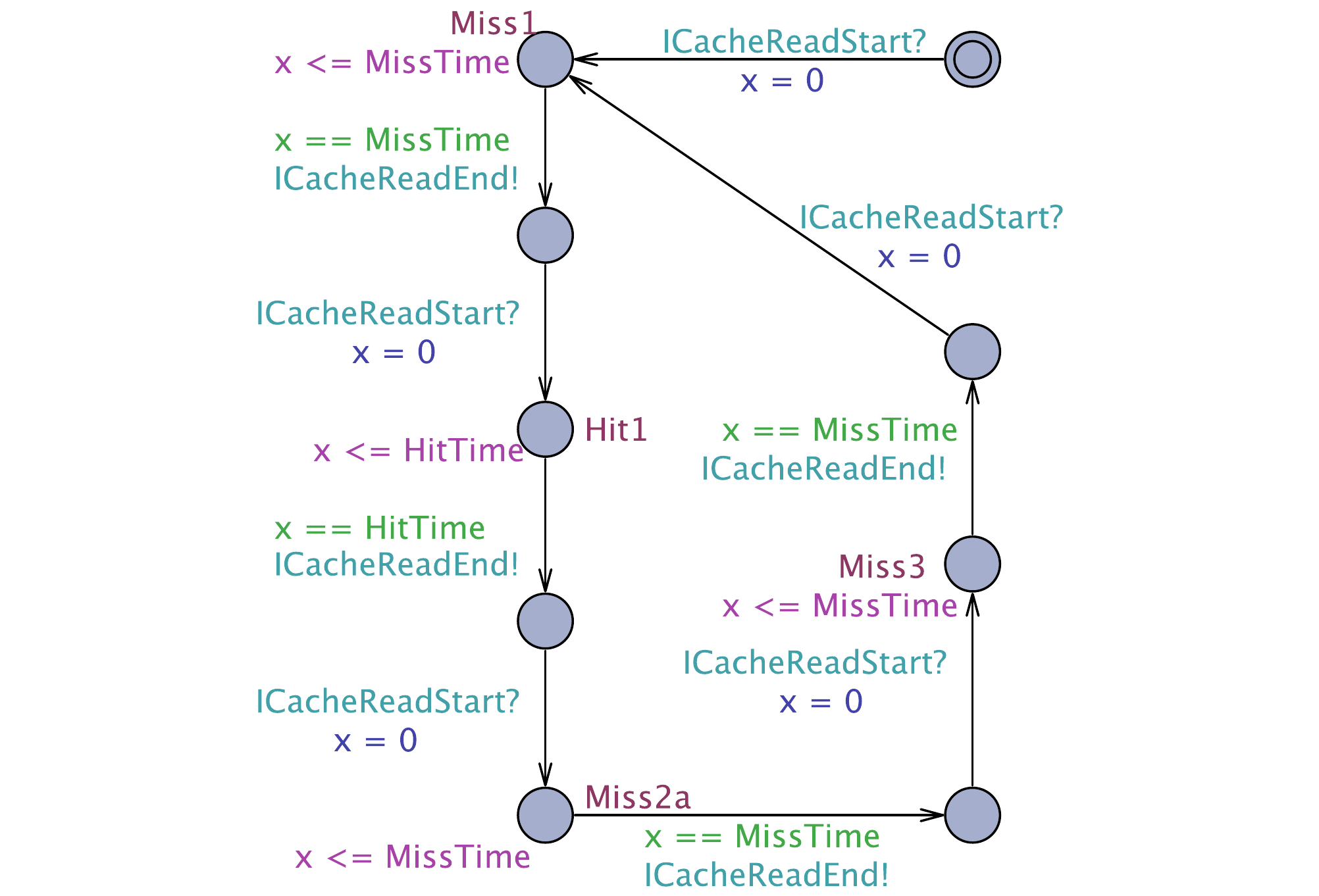}
}{%
  \caption{A small cache model}%
  \label{fig-mini-cache}
}
\capbtabbox{%
 \begin{tabular}{||c|c|c|c||}\hline
$N$ & \multicolumn{2}{c|}{States Explored} & WCET \\\cline{2-3}
&  Explicit Model & Small Model  & \\\hline
1 & 549  & 147  &  396\\\hline
2 &  1055 & 196  & 396 \\\hline
3 &  1626 &  245 & 396 \\\hline
4 &  2267 &  294 & 396 \\\hline
5 &  2953 &  343 & 396 \\\hline
6 &  3699 &  392 & 396 \\\hline
7 &  4505 &  441 & 396 \\\hline
8 &  5371 &  490 & 396 \\\hline
9 &  6297 &  539  & 396 \\\hline
10 &  7283 &  588  & 396 \\\hline
\end{tabular}
\vspace*{1cm}
}{%
\caption{States explored for computing the WCET}
\label{tab-states}
}
\end{floatrow}
\end{figure}

\paragraph{\it \bfseries Equivalent states in cache models.}
The execution time of a sequence of instructions depends on the sequence of hits and misses and the time it takes for the CPU to execute each instruction.
Assume that the time to execute each instruction is $0$ and only the cache hits and misses impact the execution time.

With a model that explicitly represents the cache, we can precisely track the hits and misses.
For instance for the sequence $\sigma_1 = 1.2.3.1$ and a cache of size $3$, the successive states of the cache are $\varnothing$, $1$, $1.2$, $1.2.3$ and we  obtain the following
sequence of pairs (instruction, Hit/Miss) : $(1,M).(2,M).(3.M).(1,H)$. This sequence  fully determines the execution time.
It turns out that  the sequence of  instructions $1.4.5.1$ would produce exactly the same execution time (if $4$ and $5$ have the same execution time as $2$ and $3$).

Now let $P_1$ be the program that comprises of two sequences of instructions $\pi_1 = 1.2.3.w$ and $\pi_2 = 1.4.5.w$, with $w= k.w'$ a finite sequence of instructions that does not contains $\{1,2,3,4,5\}$.
Assume $1$ takes $1$ cycle to execute,  $2$ and $3$ both take $2$ cycles and $4$ and $5$ both take $1$ cycle to execute.
A cache hit costs $1$ and a cache miss $10$.
If we want to compute the WCET of $P_1$ we have to analyse both $\pi_1$ and $\pi_2$.
As $w$ is the same in $\pi_1$ and $\pi_2$, the \texttt{PC} of the program before $w$ is $k$ and  is the same after $1.2.3$ and $1.4.5$.
But the cache is in a different state after $1.2.3$ and $1.4.5$.
After $1.2.3$ the current time given by \texttt{GBL\_CLK} is $35$ and after $1.4.5$, \texttt{GBL\_CLK} is $33$.
To compute the WCET in our model-checking based approach with explicit cache, we enumerate the (symbolic) state space of the synchronised product of the
program and the cache. 
In the state space we can reach two different symbolic states $s_1 = (k, 1.2.3, \texttt{GBL\_CLK}=35)$ and $s_2 = (k, 1.4.5, \texttt{GBL\_CLK}=33)$.
They have to be explored (\eg in a Depth-First Search) to compute the WCET.
From $s_1$ and $s_2$ the sequences of pairs $(i,\alpha)$ of (instructions, Hit/Miss) will be the same when executing $w$. 
As we know that \texttt{GBL\_CLK} is the largest after $1.2.3$, is never reset, we do not need to explore the state $s_2$ and the full path $\pi_2$ because it will surely yield a smaller execution time.
The states $s_1$ and $s_2$ can be considered as equivalent  \emph{given} the program we analyse: they
will generate the same sequences of Hits and Misses and the same execution time.

Computing equivalent cache states can be achieved using abstraction refinement techniques as described in~Section~\ref{sec-cache}. 
The following example shows that using an abstract cache can lead to drastic reductions in the state space of the system to analyse while presering the WCET.

\paragraph{\it \bfseries An example.}
Consider  the program of Fig.~\ref{fig-prog-auto}.
We fix the maximum number of iterations to $M=5$.
We can compute the WCET of the program with an explicit cache of size $2$.
A model in which equivalent cache states are collapsed is given by Fig.~\ref{fig-mini-cache}. 
With a cache of size $2$ all the runs in the program generates sequences of Hit and Miss of the form $(M.H.M.M)^*$.
Hence the automaton of Fig.~\ref{fig-mini-cache} is a good cache model for this program for any number of switches $N$.

Table~\ref{tab-states} gives the number of states explored\footnote{We used \uppaal and the command \texttt{verifyta -t0 -u simple-prog.xml.}} to compute the WCET with the explicit cache and
the small cache of Fig.~\ref{fig-mini-cache}.
As can be seen, many states of the cache are equivalent and this is captured by the small model whereas the explicit model generates all the configurations.

The explicit state cache model can be used with any program to analyse but may inflate the state space to be explored for computing the WCET.
Another disadvantage of the explicit state cache model is that the initial state has to be fixed.
In the next section we describe how to compute small cache models \emph{for a given program}.

%!TEX root = mars-ws.tex

\section{Computing abstract cache models}
\label{sec-cache}

As demonstrated in the previous section, using an explicit cache model can be detrimental to the model-checking approach as many equivalent (time-wise) cache states may be explored.
The objective of this section is to indicate how we can compute small cache models that are good abstractions \emph{for a given program}.
To do this, we use an abstraction refinement technique introduced in~\cite{HeizmannHP-sas-09,HeizmannHP-cav-13}.

\begin{figure}[thbp]
\centering
\begin{tikzpicture}[scale=1,node distance=2cm and 2cm, very thick, bend angle=20,bend angle=10]
\tikzset{green/.style={rectangle, rounded corners,thick,draw=black},
    red/.style={rectangle, rounded corners,thick,draw=black},
    adam/.style={rectangle, rounded corners,thick,draw=black},
    module/.style={rectangle, rounded corners,thick,draw=black}}
 \node[module](0,0) (init) {\stack{Compute the WCET $w$  \\  in $Aut(P) \times Aut(C)$}};
 \node[module,right of=init, xshift=4.8cm] (checkfeas) {\stack{$t$ feasible?}};
  % \node[below of=init,green] (nobug) {No error in $p$};
  \node[right of=checkfeas,red,xshift=1.4cm] (bug) {\stack{WCET is $w$\\ $t$ is a witness}};
  \node[above of=init,xshift=-2.8cm] (start) {$Aut(C) = HitOrMiss$};
\node[above of=init,xshift=3.5cm,red] (refine) {\stack{Compute refinement $O(t)$ \\
    $Aut(C) := Aut(C) \setminus O(t)$}};

\path[->] 
  (init) edge[swap] node[yshift=.6cm] {\stack{Get a witness trace $t$ \\ of duration $w$}} (checkfeas)
  (checkfeas) edge[draw=black] node {{Yes}} (bug)
   ;

 \draw[->] (checkfeas) |- node[xshift=.3cm,yshift=-.4cm] {No}  (refine.east);
 \draw[->] (refine.west) -| ($(init.north) + (.3cm,0)$);
  \draw[->] (start) -|  ($(init.north) + (-.3cm,0)$);

\end{tikzpicture}\caption{Trace Abstraction Refinement Algorithm for computing WCET}
\label{fig-trace-refinement}
% \vspace{-.6cm}
\end{figure}
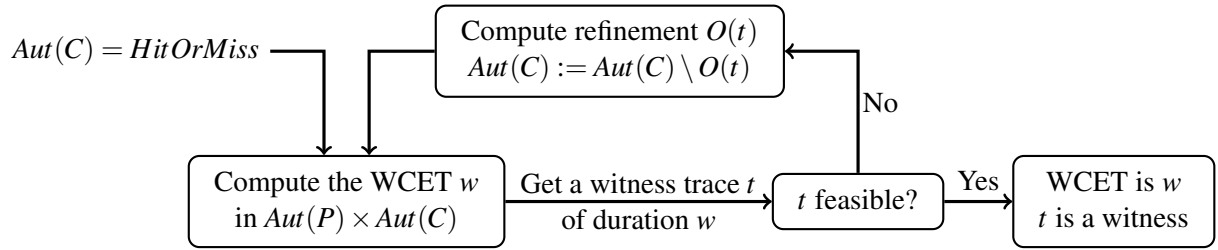
\paragraph{\it \bfseries Computation of WCET via trace refinement.}
The \emph{trace abstraction refinement technique} was originally developed to analyse imperative programs.
We adapt here it  to compute the WCET of a program. 
Our  method works as  follows to compute the WCET of program $P$:
\begin{enumerate}
  \item start with the most abstract model of the cache: every cache access can be either a hit or a miss;
  \item compute the WCET, $w$, of $P$ with the current model of the cache.
  \item  get a witness trace $t$ (sequence of pairs (instructions, Hit/Miss)) that 
    yields this execution time $w$.
  \item check if $t$ is \emph{feasible} in a concrete model of the cache. $t$ contains the sequence of instructions
    and using the specification  of the cache we can compute whether the corresponding sequence
    of hit and miss is feasible or not. 
    Notice that we do not need to specify an initial state for the cache but rather check that there is an initial state of the cache such that $t$ is feasible.
  \item if $t$ is feasible, the WCET is $w$ and $t$ is a witness trace.
  \item otherwise $t$ is infeasible in any concrete model of the cache, and we have to refine the abstract cache. 
    The refined abstract cache  model should not allow $t$. We iterate this process and 
    re-start at step~2.
\end{enumerate}
Step~6 above is the instrumental step of the trace abstraction refinement method: from one infeasible trace, we can compute a \emph{set} of traces that are infeasible for the same reason. From $t$ we can obtain a regular language of traces that are infeasible and represent it by a finite automaton $O(t)$. 
This crucial step is based on the computation of \emph{interpolants}. We have defined a logic for caches and the corresponding notion of interpolants that enables us to compute $O(t)$ for each infeasible trace $t$.

Fig.~\ref{fig-trace-refinement} gives an algorithm to compute the WCET of a program $P$ 
by iteratively refining the initial abstract model of the cache.
% using a trace abstraction refinement loop is given in Fig.~\ref{fig-trace-refinement}. 
The initial abstract model for the cache is the timed automaton $HitOrMiss$ 
% (appendix~\ref{app-hitMissAuto}) 
which allows the cache to generate a Hit or a Miss for each memory  access.
When an infeasible trace is discovered, we refine the cache model by removing a set of infeasible traces
given by $O(t)$.
Notice that the cache model we iteratively compute does not make any assumption on the initial state of the cache. 
The traces captured by the automata $O(t)$ are not feasible for any initial state of the cache.

%!TEX root = mars-ws.tex

\section{Conclusion}
We have presented a method to compute the WCET based on timed automata and  real-time model-checking with \uppaal.
The method we propose computes the model of the instruction cache as needed by refining the initial abstract model of the cache.
This yields  a reduction in the search space, which results in smaller models to be analysed and is key step towards the scalability of our timed automata based  method.
Another interesting feature of this approach is that we do not need to assume that the initial state of the cache is known. Indeed, checking whether a trace is feasible can be encoded as a SAT problem. The refined models of the cache we compute generate the cache behaviours that are possible from  \emph{some} initial state of the cache.
We are currently implementing the refinement loop  and the computation of the refined models of the cache.

\bibliographystyle{eptcs}

\vspace{-.19cm}

% \bibliography{wcet}
% \bibliography{wcet,biblio-multi-thread}

\newpage

\appendix

%!TEX root = mars-ws.tex

\section{Instruction cache \uppaal code (declarations)}
\label{app-cache-full}
\begin{lstlisting}[language=C]
int MissTime = 20;             // number of cycles taken by a Miss
int HitTime = 2;               // number of cycles taken by a Hit

const int cache_size = 2;      // cache size is 2
const int bot = - 1;           // dummy place holder element

int delayTime;                 // how long it takes to fetch the data

int cache_content[cache_size]; // array to hold the cache content

clock x;

// initialise array to bot
void initCache() {
  int i;
  for (i = 0; i <= cache_size; i++) {
    cache_content[i] = bot;  // initialised to bot means not in
  }
}

// check if element is the cache
int find(int element){
  int i;
  // scan cache 
  for ( i = 0; i <= cache_size; i++) {
    if(cache_content[i] == element) return i;
  }
  return -1;  //  not found
}

void insert(int k,int element){
  int i;
  //  maintain the cache in FIFO order
  for ( i = k; i >= 1; i--) {
     cache_content[i] = cache_content[i-1];
  }
  cache_content[0] = element;
}

void access(int thePC){
  int i = find(thePC);  // is it in?
  if(i <= 0) { 
    // not found, cache Miss, insert
    insert(cache_size - 1, thePC);
  } else { 
    // found, cache Hit, update FIFO cache
    cache_content[i] = bot;
    insert(i, thePC);
  }
}
\end{lstlisting}

\end{document}